\documentclass[12pt]{article}

\usepackage{graphicx}
\usepackage{subfig}

\usepackage{placeins}

\usepackage{authblk}
\usepackage{amsmath}
\usepackage[margin=1in]{geometry}

\usepackage[math]{cellspace}
\begin{document}

\title {Comment on ``Photons can tell `contradictory' answer about where they
have been''}

\author[1]{Gregory Reznik}

\affil[1]{\small Raymond and Beverly Sackler School of Physics and Astronomy, Tel-Aviv University, Tel-Aviv 69978, Israel}
\affil[2]{Max-Planck-Institut f\"{u}r Quantenoptik, Hans-Kopfermann-Stra{\ss}e 1, 85748 Garching, Germany}
\affil[3]{Department f\"{u}r Physik, Ludwig-Maximilians-Universit\"{a}t, 80797 M\"{u}nchen, Germany}

\author[2,3]{Carlotta Versmold}

\author[2,3]{Jan Dziewior}
\affil[4]{Schmid College of Science and Technology, Chapman University, Orange, California 92866, USA}
\affil[5]{Institute for Quantum Studies, Chapman University, Orange, California 92866, USA}

\author[2,3]{Florian Huber}

\author[2,3]{Harald Weinfurter}

\author[4,5]{Justin Dressel}

\author[1,2,3,5]{Lev Vaidman}

\date{}

\maketitle

\begin{abstract}
Yuan and Feng [Eur. Phys. J. Plus 138:70, 2023] recently proposed a modification of the nested Mach–Zehnder interferometer experiment performed by Danan et al. [Phys. Rev. Lett. 111:240402, 2013] and argued that photons give ``contradictory'' answers about where they have been, when traces are locally imprinted on them in different ways.
They concluded that their results are comprehensible from what they call the ``three-path interference viewpoint'', but difficult to explain from the ``discontinuous trajectory'' viewpoint advocated by Danan et al.
We argue that the weak trace approach (the basis of the ``discontinuous trajectory'' viewpoint) provides a consistent explanation of the Yuan-Feng experiment.
The contradictory messages of the photons are just another example of photons lying about where they have been when the experimental method of Danan et al.~is applied in an inappropriate setup.

\end{abstract}

%\maketitle

%--------------------------------------------------------------------------

Recently, Yuan and Feng (YF) \cite{YF} considered a new modification of the nested Mach-Zehnder interferometer (MZI) experiment \cite{Danan}.
They argued that this example presents a new difficulty for tracing the travel history of photons by the “weak measurement” method. 
They write:

\begin{quote}
    In the nested MZI, every mirror vibrates in distinguishable frequencies which can be read out from output signal by Fourier analysis. The presence or absence of a certain frequency is claimed to be criterion of whether a photon has reflected off the related mirror.
\end{quote}
We argue that this statement misinterprets the main point of \cite{Danan}.
Indeed, in this experiment the presence of the vibration frequency of a mirror was used as a signature of the presence of the photon near this mirror, but it was not \textit{claimed to be the criterion} for presence.
Instead, as defined in \cite{past}, the actual criterion for the weak trace analysis (WTA) is the creation of local traces, i.e., changes in the local environment such as momentum kicks of the mirrors.
Since in practice those are difficult to observe, the signal imprinted on the traveling particles, i.e., the change of the transversal degree of freedom of the photon, serves as an indirect, experimentally friendly method to witness those traces.
However, this requires that this imprint, once locally created, remains undisturbed until the particle is detected.
%The criterion for the presence was the creation of a local trace.
%In experiment \cite{Danan} the local trace was the change of the transversal degree of freedom of the photon that remained undisturbed until the photon reached the detector. 
%The observed frequencies were witnesses of local traces, which were defined as the criterion for the presence in \cite{past}.

The YF example builds on the modification of the Danan et al.~experiment by Alonso and Jordan \cite{AJDove} who pointed out that introducing the Dove prism inside the inner interferometer leads to the appearance of the frequency of the mirror near which no significant trace was created.
The Dove prism disturbs the transversal degree of freedom of the photon, such that the detected photons no longer provide faithful information about the local traces \cite{VT}.

\begin{figure}
\centerline{\includegraphics[width=1\textwidth]{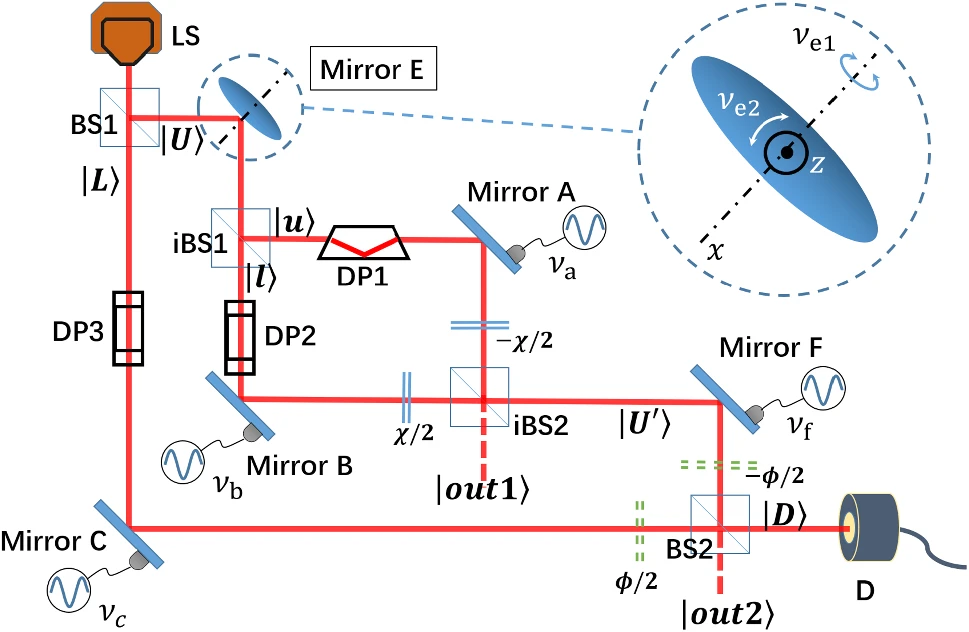}}
\caption{ {\bf Yuan and Feng modification of Danan et al. experiment with Dove prism. (Fig.1 of \cite{YF}).}
The interferometer considered in two situations. The inner MZI is tuned either to constructive or destructive interference towards mirror $F$.
In both cases, the mirror $E$ is oscillating around the $z$ and $x$ axes with different frequencies.
And in both cases only one frequency is observed at the detector.
Our explanation is that the photon was present near mirror $E$ in the setup with constructive interference and was not present in the setup with destructive interference.
Presence and absence of the frequency of vibration around the $x$ axis, which results in the modulation of the intensity at $E$, corresponds to the presence and absence of the photon there.
The presence of the photon is not shown in the signal, i.e., the frequency of the oscillation around the $z$ axis does not appear due to filtering of the signal by the inner interferometer with the Dove prism.
Changing the phase of the inner MZI by $\pi$, leads instead to filtering out the undisturbed component of the photon keeping only the orthogonal component which appears due to oscillation around the $z$ axis.
This results in a strong signal in spite of only a tiny (secondary) presence at $E$.
} 
\label{fig:1}
\end{figure}

Apart from adding the Dove prism, YF suggest two more modifications to the nested MZI experiment.
First, for mirror $E$, near which the presence of a photon is in the dispute, they consider not only a vibration around the $z$ axis, but also around the $x$ axis with different frequencies, see Fig. \ref{fig:1}.
The second modification (which will be discussed later) is to consider the nested MZI with the Dove prism when it is tuned to constructive interference in all its parts.
In both cases, they found that the photons tell a contradictory story about their presence in the vicinity of the mirror $E$.

For an ideal mirror a vibration around the $x$ axis creates no trace on the photon, however, YF consider a non-ideal mirror with reflectivity varying over the its surface.
For this mirror, the vibration is equivalent to introducing a modulation of amplitude, instead of a modulation of transversal momentum.
%Modulation of the amplitude is supposed to change the probability of postselection directly.
%YF write: ``...the amplitude of the reflected beam change periodically, which can be probed by the photon detector(D)''.
%Note that in the Danan et al. experiment the signal was the difference of the intensity on the upper and lower part of the detector and not the total intensity, so different analysis of the output should be done to observe the signal caused by the vibration around the $x$ axis. 
%\section{Destructive Interference}
In the setup with the Dove prism, in which the inner interferometer is tuned to destructive interference towards $F$, the frequency of the vibration around the $x$ axis does not appear in the observation of the total intensity of the detected photons.
On the other hand, we do see the frequency of the vibration around the $z$ axis in the modulation of the detector signal.
%This represents a ''contradictory answer'' according to YF.
%This represents the first ''contradictory answer'' discussed by YF.
In the words of YF the ``photons tell `contradictory' answer about where they have been.''
%They concluded that their results are comprehensible from the three-path interference viewpoint, but difficult to explain from the ``discontinuous trajectory'' viewpoint advocated by Danan et al. 

%This apparent contradiction is addressed/resolved by the explanation given by Vaidman and Tsutsui (VT) \cite{VT} as a response to Alonso and Jordan \cite{AJDove}.
%Although YF cite \cite{VT}, they do not respond to this point.
The explanation of a similar contradiction has already been provided by Vaidman and Tsutsui (VT) \cite{VT} in their response to Alonso and Jordan \cite{AJDove}.
YF cite VT, but apparently do not accept that explanation. % given there. %Vaidman Tsutsui (VT) paper \cite{VT}
Indeed, in the last paragraph of Section 2 of \cite{VT} it is mentioned that, in a version of the experiment with a single Dove prism, the frequency of vibration of the mirror $E$ around the $y$ axis does not appear in the signal, which is very similar to the claim of YF.
This absence of the signal has a clear explanation according to the WTA, namely that the photons were not present near the mirror $E$.
%The frequency of the vibration of the mirror $E$ around the $x$ or $y$ axes does not appear in the signal, because the postselected photons were not present near the mirror $E$. 
More precisely, the photons did not have a primary presence at $E$, which is defined as leaving a trace of the order of the trace that a well-localized photon would leave, (though the photon had a ``secondary presence'' at $E$ \cite{vaidman2014tracing}).

VT also explained why, surprisingly, there {\it is} a strong signal of the frequency of  vibration around the $z$ axis, even though the photons are not present at $E$ according to the WTA.
%It is more difficult to explain why there {\it was} a strong signal of frequency of the vibration around $z$ axis.
Vibrations around the $y$ and $z$ axes leave similar traces on the photon, but VT showed that the nested MZI with the Dove prism placed in the x-y plane leads to a special interference effect for the postselected photons which amplifies the signal of the vibration around the $z$ axis.
(A rotation of the Dove prism by 90 degrees around the photon propagation direction would lead to amplification of the other frequency.) 

Let us present the explanation in some detail. The mirror introduces a small change in the quantum state of the photon with
%What happens is that the mirror introduces a small change in the quantum state of the photon
\begin{equation}\label{mirror}
|\chi \rangle \rightarrow|\chi^\prime \rangle \equiv \eta\left( |\chi \rangle + \epsilon |\chi^\perp \rangle \right),
\end{equation}
where $| \chi \rangle$ is the quantum state of the photon when the mirror is not rotated, $| \chi^\perp \rangle$ denotes the component of the modified state $| \chi^\prime \rangle$ that is orthogonal to $| \chi \rangle$, $\epsilon \ll 1$, and $\eta$ is the proper normalization.
A vibration of the mirror around the $z$ axis causes oscillations of the amplitude of the orthogonal component which lead to the observed signal.
The inner MZI filters out the undisturbed photon wave $| \chi \rangle$  from the (dark) output port which leads towards mirror $F$, but keeps the component $| \chi^\perp \rangle$ due to constructive interference.
This leads to an amplification effect such that the tiny secondary presence near $E$ provides the same signal as the large primary presence in other mirrors (e.g. $A$) vibrating in a similar manner.

%\section{Constructive Interference}

The second novel point of YF was the consideration of the nested MZI tuned to constructive interference toward the final detector.
In this case, the WTA tells us that the photon was present at $E$.
Accordingly, the frequency of oscillations around the $x$ axis is observed.
%We do observe the frequency of oscillations around the $x$ axis.
This is not surprising.
A modulation of the absorption leads to a modulation of the probability of reaching the final detector.
However, we do not see (in the first order) the frequency of vibration around the $z$ axis in the signal.
The reason is that now the inner MZI (with the $\pi$ shift in the phase) filters $| \chi^\perp \rangle$.
Thus, due to the prism, destructive interference towards $F$ suppresses this particular signal.
%Due to the Dove prism, it makes destructive interference toward $F$.
Of course, it does not filter the modulation of the total amplitude caused by the vibration around the $x$ axis.

%\section{Presence from Carried Information (about local properties)}

Our first answer to YF is that according to the WTA the criterion of presence is a local trace which is faithfully transferred to the detector in the original Danan et al.~setup, but spoiled by introducing the Dove prism.
In the case with destructive interference it provides an amplification leading to a strong signal from photons which are hardly present, and in the case with constructive interference it suppresses the signal from present photons.
Thus, the criticism by YF is invalidated by showing that their proposed experimental setup is in fact not suitable to observe local traces, which are the actual \textit{criterion} of the WTA.

However, the presence of quantum particles in the past can also be approached from a different conceptual point of view, distinct from the WTA, as recently formulated in \cite{Bhati}:
%However, the evidence for the presence at different mirrors in nested MZI experiments can be viewed in a conceptually different way:
``The photon was in a particular location if it carries information about some local properties of this location.''
This corresponds to YF writing:

\begin{quote}
    If the presence or absence of a certain frequency was the criterion of whether the photon had reflected off the related mirror, then contradictory results would be obtained: The emerging frequency indicated that the photon had reflected off the mirror, whereas the vanishing one indicated that the photon had never reached the mirror! For this reason, we conclude that the presence of one frequency can be regarded as a sufficient condition of the fact that the photon has reflected off the related mirror. But, if the frequency is absent, we cannot affirm that the photon does not reach the related mirror.
\end{quote}
We disagree with YF also with respect to this alternative approach to particle presence.
%We argue that 
Neither the presence nor the absence of the photon near the mirror can be inferred \textit{just} from the appearance or disappearance of the frequency of the mirror in the Fourier analysis of the signal. %obtained from the photons.
We should also take into account the efficiency of the local interaction that introduces changes in the state of the photon reaching the detector. 
The signal in the experiment has to be compared with the signal in a similar setup where the photon was forced to be near $E$.
For the MZI tuned to destructive interference, this has been done in detail in \cite{AswerBhati}.
For the case of tuning to constructive interference, one can apply the analysis of YF.
Even in the case where the photon passes only through $E$ (and so there is no doubt that it was present in $E$), still the frequency of the oscillation around the $z$ axis does not appear in the signal.
Localization at $E$ would only remove the contribution from the path $C$ from Eqs.~(13) and (14) of YF but would not lead to an appearance of $\varphi_e$, corresponding to information about the oscillation of mirror $E$ around the $z$ axis.

In their conclusions, YF write: ``These results are comprehensible in the three-path interference viewpoint, but difficult to explain in the `discontinuous trajectory' viewpoint''.
Let us also comment on the ``three-path interference viewpoint''.
%Indeed, the results of the experiments, the behavior of the detector in the output port of the nested MZI are explained by standard quantum mechanical calculations (called 'three-path interference viewpoint' by YF) just as given in Supplement IV of \cite{Danan}.
Indeed, the results of the experiments, i.e., the signals at the detector in the output port of the nested MZI are explained by the ``three-path interference''.
In our view, this is just the standard quantum mechanical explanation, very much the same as the one presented in Supplement IV of \cite{Danan}.
%This obviously holds true for all predictions of the TSVF and the WTA which provide an alternative \textit{formalism} especially suitable to the analysis of pre- and postselected scenarios.
%Furthermore, crucially the WTA introduces a criterion for past particle presence which is not present in the standard quantum mechanical formulation.
However, we could not see an answer to the question ``Where was the photon inside the interferometer?'' in YF and in other papers presenting or adopting the ``three-path interference'' approach \cite{YuanThreePath,YuanPhotonsHide}. %we could not see a criterion for particle presence, a counterpart to the weak trace,
From their discussion of the signal from the mirror $E$, it seems that according to their picture the photon was present near $E$ also when the inner interferometer was tuned to destructive interference towards $F$.
But according to the classical way of thinking about particle presence, the photons pass through $C$ only \cite{Englert} (see also \cite{Peleg,EnglertReply}).

In summary, the results predicted by YF have a consistent explanation both in the WTA and when using an approach which is fundamentally based on the information carried by the particles.
Both of these approaches to particle presence predict exactly the same discontinuous trajectories in the case of tuning to destructive interference \cite{AswerBhati}.
The work of YF neither shows an inconsistency of the WTA nor does it introduce a new concept of presence which would lead to conclusions differing from the WTA.
%The main result of our Comment is that the YF results do have consistent explanation in the weak-trace approach (which provides a discontinuous trajectory in the first YF setup).

\section*{}
This work has been supported in part by the National Science Foundation Grant No. 1915015 and the U.S.-Israel Binational Science Foundation Grant No.~735/18 and by the Israel Science Foundation Grant No.~2064/19.
Furthermore, we want to acknowledge support by the DFG both under the project Beethoven 2 (2016/23/G/ST2/04273, 38144572) and under Germany's Excellence Strategy EXC-2111 390814868.

%unsrt
\bibliographystyle{unsrt}

\end{document}